\documentclass[prl,twocolumn]{revtex4}
\usepackage{graphicx}
\newcommand{\pnas}{Proc. Natl. Acad. Sci. USA}
\newcommand{\jpcb}{J. Phys. Chem. B}
\newcommand{\caltech}{Norman Bridge Laboratory of Physics 12-33, California Institute
of Technology, Pasadena, CA 91125}

\begin{document}
\title{Photon statistics and dynamics of Fluorescence Resonance Energy Transfer}
\author{Andrew J. Berglund}
\email{berglund@caltech.edu}
\author{Andrew C. Doherty}
\author{Hideo Mabuchi}
\affiliation{\caltech}
\begin{abstract}
We report high time-resolution measurements of photon statistics
from pairs of dye molecules coupled by fluorescence resonance
energy transfer (FRET). In addition to quantum-optical photon
antibunching, we observe photon bunching on a timescale of several
nanoseconds. We show by numerical simulation that configuration
fluctuations in the coupled fluorophore system could account for
minor deviations of our data from predictions of basic F\"{o}rster
theory. With further characterization we believe that FRET photon
statistics could provide a unique tool for studying DNA mechanics
on timescales from $10^{-9}-10^{-3}$ s.
\end{abstract}
 \maketitle

Fluorescence Resonance Energy Transfer (FRET) has become a
widespread tool for probing molecular structure and dynamics.
Recent demonstrations of single-molecule sensitivity in optical
assays based on FRET \cite{Deniz99a,Ha99b} have led to significant
advances in our understanding of topics such as RNA folding and
ribozyme function \cite{Weiss00a}. Interpretation of FRET
data generally relies on a simple physical model
involving near-field dipole-dipole interactions between molecules,
which was first proposed by F\"{o}rster \cite{Forster}. While some
basic features of this model pertaining to steady-state solutions
have previously been verified experimentally, dynamical details
have been largely inaccessible. In this Letter,  we report the use
of a Hanbury-Brown Twiss apparatus to record photon statistics of
the light emitted by FRET-coupled dye pairs with nanosecond
resolution, and show that a careful comparison of our data with
predictions of F\"{o}rster theory supports the basic model but
indicates a class of additional factors that must be considered.
Our analysis suggests that conformational fluctuations of the
substrate for the FRET-coupled dyes could be such a factor, which
in turn points to the intriguing possibility of utilizing FRET
photon statistics for novel assays in DNA and protein mechanics.

\begin{figure}[b]
\scalebox{0.3}[0.3]{\includegraphics{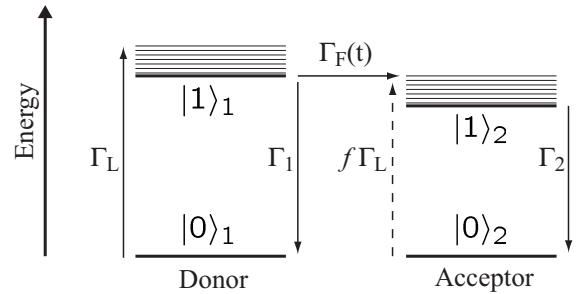}}
\caption{\label{fig:levels} Energy-level diagram of molecular
states relevant to the basic F\"{o}rster model. Donor states on
the left are coupled by FRET to acceptor states on the right.}
\end{figure}

The FRET process involves non-radiative transfer of energy from a
donor, which absorbs a photon of incident light, to an acceptor
that is not directly coupled to the incident light. Detection of
acceptor fluorescence is thus a simple indicator of FRET activity.
A schematic energy-level diagram of molecular states is shown in
Fig.\ \ref{fig:levels}. Under appropriate conditions of spectral
overlap, F\"{o}rster theory \cite{Forster} predicts that the rate
$\Gamma_F$ of energy transfer varies as $\Gamma_F\propto
\kappa^2/R^6$, where $\kappa^2$ depends on the orientation of the
fluorescent species and $R$ is the distance between them. For
commonly-used organic dyes, the sensitivity of $\Gamma_F$ to
variations in $R$ is greatest in the range of several nm; hence
experimental determination of $\Gamma_F$ yields information on
distance scales relevant to biological macromolecules. The
$R^{-6}$ distance dependence of F\"{o}rster theory has been
experimentally confirmed using `ruler' strands of DNA
\cite{Deniz99a}.

Intensity correlation functions of the light emitted by FRET-coupled dyes pairs carries further information about the molecular
physics of FRET. The second-order intensity correlation function
(for a {\em stationary} process) gives the normalized time-average
intensity $I_k(t+\tau)$ of mode $k$ at time $t+\tau$ multiplied by
the intensity in mode $j$ at time $t$:
\begin{equation}\label{g2}
g_{jk}^{(2)}(\tau) = \frac{\langle I_j(t)
I_k(t+\tau)\rangle}{\langle I_j(t)\rangle \langle I_k(t)\rangle}.
\end{equation}
For $j=k$, this quantity is the autocorrelation of the
intensity field $j$. Nonclassical photon statistics
\cite{Mandel95a}, for which $g^{(2)}_{jj}(\tau)> g^{(2)}_{jj}(0)$
for some $\tau>0$, were first observed in an atomic beam
\cite{Kimble77a}. Since then, photon antibunching has been
observed in a variety of systems 
\cite{Kask85a,Diedrich87a,Michler00a, Fleury00a}. 

We have measured $g^{(2)}_{jk}(\tau)$ for individual FRET-coupled
Cy3 and Cy5 dye molecules tethered to DNA \footnote{Integrated DNA
Technologies molecular beacons.}. We show that our data should be
sensitive not only to the mean values of FRET parameters, but also
to underlying molecular processes that perturb their values on any
timescale down to that of the radiative lifetimes. The techniques
presented here thus provide potential experimental access to
molecular dynamics that influence radiative level structures and
couplings in the range 10-1000 ns, in a manner complementary to
that of established techniques such as fluorescence correlation
spectroscopy. Intriguing examples of such dynamics include
photochemical processes, chemical shifts arising from changes in
the local environment, single base-pair fluctuations in DNA
secondary structure formation, and conformational fluctuations on
much shorter timescales than have previously been studied using
FRET. Molecular dynamics simulations of nucleic acid mechanics are
generally tractable only for integration times $\sim 10$ ns
\cite{Cheatham00a}, so that experimental access to these
timescales may provide fruitful contact between theory and
experiment.

\begin{figure}[t]
\scalebox{0.2}[0.2]{\includegraphics{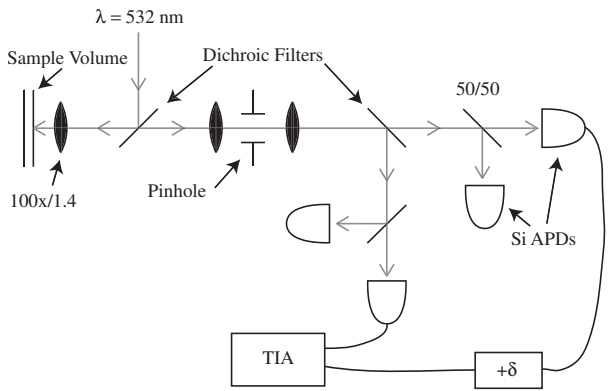}}
\caption{\label{apparatus}Schematic diagram of the apparatus. When
making cross-correlation measurements, the $50/50$ beam-splitters
are removed to improve collection efficiency. Spectral filters and
focusing optics at the APDs are not shown.}
\end{figure}

A diagram of our apparatus appears in Fig.\ \ref{apparatus}. It
consists of confocal imaging optics coupled to Hanbury Brown-Twiss
(HBT) detection channels \cite{HanburyBrown56a}. We
focus $140$ $\mu$W of 532 nm laser light between glass cover
slips through a diffraction-limited microscope objective (Carl
Zeiss). Fluorescence is collected by the same
objective and imaged onto a 100 $\mu$m-diameter pinhole. A dichroic filter
separates Stokes-shifted fluorescence light from scattered
excitation light. A second dichroic filter separates donor
fluorescence ($~570$ nm) and acceptor fluorescence
($~670$ nm) into separate HBT channels each
containing a 50/50 beam-splitter, spectral filters and 2 avalanche photodiode single-photon counters (APDs). In each experiment,
photon arrival times at one pair of detectors are recorded with
sub-ns resolution by a time-interval analyzer (TIA). An electronic
delay of $\delta=50$ ns is imposed in one channel to avoid small
time-interval crosstalk in the TIA.

We monitor fluorescence from dual-labelled DNA hairpins in aqueous
buffer at room temperature. The donor and acceptor are
tethered at complementary positions, so that they exhibit a high
FRET efficiency. In a typical experiment, we place $1 \mu
$L of 1 nM dye-labelled DNA solution between the cover
slips. The axial position of the microscope objective is actively
locked by a piezo-electric translator so that it is
stable to $\lesssim 100$ nm for periods much longer than a typical
experimental run ($\sim6$ hrs). We choose a low enough DNA
concentration that there are no molecules in the imaging volume
for a large fraction of the observation time. The count rate at
the detectors then shows background light punctuated by
bursts of fluorescence as individual DNA strands diffuse
through the imaging volume. Since fluorescence from separate DNA
strands is {\em uncorrelated}, the presence of multiple molecules
in the imaging volume reduces the sharpness of features in the
measured correlation functions.

\begin{figure}[t]
\scalebox{0.5}[0.5]{\includegraphics{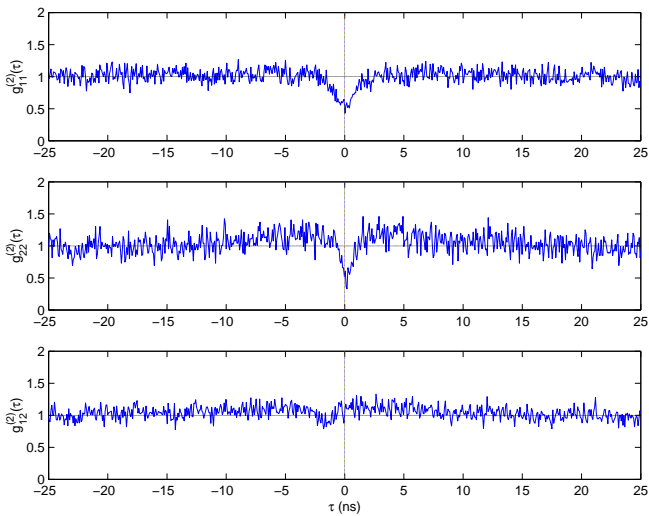}}
\caption{\label{corr_exp} Measured correlation functions for high
FRET-efficiency donor-acceptor pair fluorescence. Top: donor
intensity autocorrelation (partially contaminated by inactive
Cy5). Middle: acceptor intensity autocorrelation. Bottom:
donor-acceptor cross-correlation, the average intensity in the
acceptor channel, given a photon arrival in the donor channel at
$\tau = 0$. See text for a discussion of the bunching at $\sim
\pm5$ ns in the acceptor autocorrelation.}
\end{figure}

To measure $g_{jk}^{(2)}(\tau)$, we choose a pair of detectors and
histogram the time interval between photon arrivals, keeping only
data for which both detection channels were active (the
TIA-limited channel dead time is $243$ ns). We choose a
threshold count rate for each channel (the maximum expected count
rate over a 1 ms interval, given the mean count rate over the
$\sim 1$s local measurement interval), and keep only
those data for which at least one channel exceeds this threshold.
In this way, we reduce the contribution from background light
recorded when no molecule is present in the imaging volume. Measured correlation functions, averaged over many molecular
transits, are shown in Fig.\ \ref{corr_exp}. The precise time delay imposed by optical and electronic path differences (the $\tau=0$ point) is determined by measuring correlation functions of a pulsed LED.

Interpretation of the experimental results proceeds from a
straightforward model for Monte Carlo simulation of this and
similar experiments. The donor and acceptor are organic molecules
attached to a complex substrate. Electronic excitation
is followed by fast rotational and vibrational relaxation, so our model assumes
negligible coherence between electronic states. We represent the
donor and acceptor, labelled $j=1,2$ respectively, as two-level
emitters with basis states $\{|0\rangle_j,|1\rangle_j\}$ (see
Fig.\ \ref{fig:levels}) and lowering operator $\sigma_j$. Since we assume
no coherent interactions, we write a
master equation for the time evolution of the density operator of
the system $\rho$ with only incoherent (jump) terms:
\begin{equation}\label{master_equation}
\frac{\partial\rho}{\partial t} = \sum_{m=1}^M \Gamma_m(t)
\left\{\Lambda_m\rho\Lambda_m^\dag -
\frac{1}{2}\left(\Lambda_m^\dag\Lambda_m\rho +
\rho\Lambda_m^\dag\Lambda_m \right)\right\}.
\end{equation}
The `jump operators' $\Lambda_m$ and associated (possibly
time-dependent) rates $\Gamma_m(t)$ represent the incoherent
transitions that can occur in the system. In our case we choose
$M=5$ possible transitions, but it is straightforward to
generalize the model to include more processes. These transitions
are: direct excitation of the donor with rate $\Gamma_L$;
off-resonant excitation of the acceptor with rate $f\Gamma_L$,
$f<1$; donor (acceptor) spontaneous emission with rate $\Gamma_1$
($\Gamma_2$); and FRET energy transfer from donor to acceptor
excited state with time-dependent rate $\Gamma_F(t)$. A FRET
transition is represented by $\Gamma_5(t) = \Gamma_F(t),\quad
\Lambda_5 = \sigma_1\otimes\sigma_2^\dag$. The other jump
operators are similarly defined. Eq. (\ref{master_equation}) is equivalent to a
linear system of rate equations for the ground and excited state
populations of the donor and acceptor, and can thus be solved
exactly when the time-variation of all parameters is specified
analytically. For stochastic parameter variation, we
resort to numerical simulation.

\begin{figure}[t]
\scalebox{0.5}[0.5]{\includegraphics{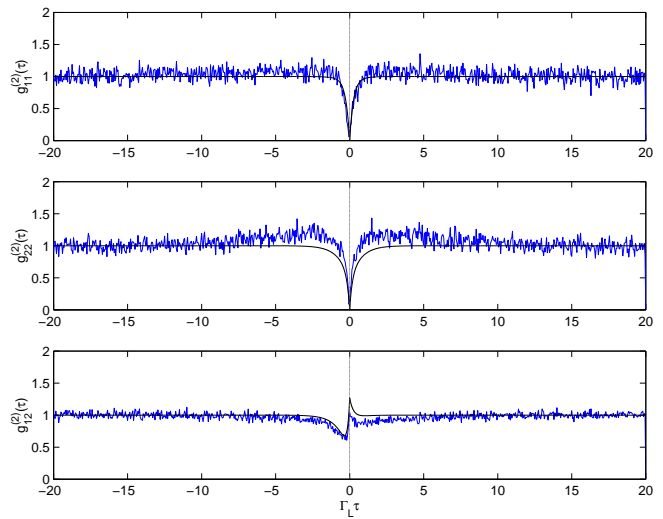}}
\caption{\label{corr_sim} Monte Carlo calculation of correlation
functions for a jump-process F\"{o}rster transfer rate with
correlation time $\tau_F=7$. Top: donor intensity autocorrelation.
Middle: acceptor intensity autocorrelation. Bottom: donor-acceptor
cross-correlation. The smooth curves are deterministic simulations
with fixed $\Gamma_F(t) = \langle\Gamma_F\rangle$. The parameters
used were $\Gamma_1=\Gamma_2=1$, $\Gamma_F^H = 10$, $\Gamma_F^L =
0$, $f=0.1$. All rates (times) are in units of the laser
excitation rate $\Gamma_L$ ($1/\Gamma_L$).}
\end{figure}

The measured autocorrelation functions in Fig.\ \ref{corr_exp}
show pronounced antibunching dips at $\tau=0$. For a {\em single}
fluorophore, $g_{jj}^{(2)}(0) = 0$ since two photons (in
the same mode) can never be emitted simultaneously, {\it i.e.}, in
a time-interval $\tau=0$. The observed value of $g^{(2)}_{jj}(0)$
is a function {\em only} of the signal-to-noise ratio
$S$ at each detector and the probability $P(N)$ that $N$
molecules are observed simultaneously. In order to understand the depth of
the $\tau=0$ minimum in the autocorrelation
functions, we make independent estimates of $P(N)$ and $S$.
Neglecting crosstalk between channels and assuming
Poisson-distributed background, it can be shown from (\ref{g2}) that
\begin{eqnarray}\label{finite_background}
g_{jj}^{(2)}( 0) &=& 1 - \sum_{N\geq 1} P(N) \frac{1}{\nu_j(N)}\\
\nu_j(N) &=& N +
\left(\frac{1}{S_j^{(1)}}+\frac{1}{S_j^{(2)}}\right) +\frac{1}{N
S_j^{(1)}S_j^{(2)}}
\end{eqnarray} where $S_j^{(n)}$ is
the signal-to-noise ratio of HBT arm $n$, mode $j$. Assuming
$P(N)$ is a Poisson distribution with mean value $\langle N \rangle$,
we determine the fraction of all photodetection events
attributed to molecular fluorescence ({\it i.e.}, exceeding the
threshold count rate criterion described above). This fraction is
$\sum_{N\geq 1}P(N)$ from which we can solve for $\langle
N\rangle$. For a typical run, we estimate $\langle N\rangle \approx 0.1$ in this
way. The probability that $N\geq 2$ molecules are observed is
therefore $<0.005$ so that we are firmly in the single-molecule
regime. We estimate the signal-to-noise ratios $S_j^{(n)}$ by
comparing the fluorescence count rate to the background count
rate. A typical value is $S \approx 5$. From signal-to-noise
and image volume occupancy statistics, we expect $g_{11}^{(2)}(0) =
0.23$ for the donor and
$g_{22}^{(2)}(0) = 0.32$ for the acceptor autocorrelation.

In our experiment, we see a large fraction ($\sim 60\%$) of low
FRET-efficiency events, indicating a subpopulation of
fluorophores exhibiting little or no FRET coupling. These events,
which we attribute to acceptor photobleaching, contribute a
background that contaminates the shape of $g_{11}^{(2)}(\tau)$.
[In this experiment, we are limited by the TIA to time-resolution
on two detectors only and are thus unable to exclude FRET-inactive
dye pairs when measuring $g_{11}^{(2)}(\tau)$.] Both the
cross-correlation and acceptor autocorrelation depend on {\em
acceptor} fluorescence events, and are therefore robust against a
bare-donor subpopulation. The depth of the observed $\tau=0$
feature in the acceptor autocorrelation is consistent with our
independent estimate based on signal-to-background and image
volume occupancy statistics.

In addition to photon antibunching on radiative timescales, we see bunching ($g^{(2)}_{22}(\tau)>1$) at longer time
intervals ($\sim 5$ ns) in the acceptor autocorrelation of Fig.\
\ref{corr_exp}. This bunching indicate clustering of
acceptor events on the same timescale, most likely arising from
fluctuations in $\Gamma_F(t)$. We expect that our fluorophores
rotate with a characteristic time of $\sim 250$ ps, so rotational
diffusion is an unlikely explanation for the observed correlations
\cite{Ha99c}. Numerical simulations investigating the influence of
rotational diffusion on $\Gamma_F(t)$ do not reproduce the
features in our data. We rather suspect FRET `intermittency',
possibly related to fast diffusion of tethered dye molecules and
their propensity to stick to DNA \cite{Edman96,Norman00}.
Intersystem crossing and spectral diffusion for fluorophores such
as Cy5 are known to exhibit longer timescales
\cite{Bernard93a,Tinn01}.

We model FRET intermittency by allowing $\Gamma_F(t)$ to jump
between a high value $\Gamma_F^H$ and a low value $\Gamma_F^L$ (as
perhaps when one or both dyes are stuck to the DNA) with a correlation time $\tau_F$.
Numerical results are shown in Fig.\ \ref{corr_sim}, where we see
antibunching followed by bunching at
$\tau_F$ \footnote{For this  ``random telegraph" 
process, the model is analytically solvable. We emphasize that the Monte Carlo techniques presented here may 
be used to simulate parameter variation with arbitrary 
statistics.}. Deterministic simulations (smooth curves) with fixed
$\Gamma_F(t)$ do not exhibit bunching in the acceptor
autocorrelation. Most calculated and observed features are
consistent with intuition based on the four-level model. Under
conditions of high FRET efficiency, donor emission rarely occurs.
However, for a sufficiently strong driving field, the excitation
rate is large compared to the acceptor emission rate, and donor
absorption may occur when the acceptor is already in its excited
state. In this ``exciton blockade" situation, FRET cannot occur
since the acceptor is already excited. Subsequent donor emission
is highly probable followed by acceptor emission a short time
later. The conditional probability for acceptor fluorescence is
therefore enhanced by observation of a donor emission event, which
is represented by a cusp at short positive times in the
cross-correlation. The dip in the cross-correlation at negative
$\tau$ can be understood in a similar way. Observation of an
acceptor photon {\it deterministically} prepares the acceptor in
its ground state. Since the 
acceptor is in its ground state, FRET occurs with high probability
since any residual donor excitation is efficiently transferred to the acceptor.
This depressed probability for donor fluorescence at short times
after acceptor fluorescence is represented by a dip at negative
$\tau$ in the cross-correlation. Cross-correlations exhibiting other types of conditional statistics have
been observed in cascaded multi-exciton emission from semiconductor quantum dots.
\cite{Regelman01a,Kiraz02a}.

 In summary, we have measured FRET photon statistics and presented an intuitive model
for interpretation of such experiments. Our data
are in basic agreement with simple F\"{o}rster theory, but the
shape of the acceptor autocorrelation is strongly suggestive of
additional dynamics at $\sim 5$ ns timescales. Numerical
simulations show that this inconsistency is resolved by the
inclusion of stochastic variations in the donor-acceptor coupling
strength. We have suggested a possible molecular mechanism for
these fluctuations, but further experiments are necessary to
characterize the biochemical details. 

Our model suggests that thermally excited bending modes of dye-labelled, rod-like molecules should be visible in $g^{(2)}_{jk}(\tau)$ for appropriate motional amplitudes and timescales. We hope to exploit this dependence in studying the conformational dynamics of semi-rigid DNA and synthetic proteins. Furthermore, future experiments incorporating direct excitation of the acceptor fluorophore will allow absolute determination of model parameters.
We believe these techniques
can be developed into an important new optical
single-molecule method for characterization of
macromolecular dynamics on nanosecond (and longer) timescales.

\begin{acknowledgments}The authors thank E.\ Winfree, R.\ Phillips,
and X.\ Zhuang for informative discussions, and T. McGarvey for
technical assistance. This research was supported by the NSF under
grant EIA-0113443 and by the W.~M.~Keck Discovery Fund. A.\ B.\
acknowledges the support of an NSF Graduate Fellowship.
\end{acknowledgments}

\end{document}